\documentclass[fleqn,usenatbib]{mnras}
\usepackage[dvipsnames]{xcolor}
\usepackage{newtxtext,newtxmath}

\usepackage[T1]{fontenc}
\usepackage{ae,aecompl}
\usepackage{times}
\usepackage{graphicx}	% Including figure files
\usepackage{amsmath}	% Advanced maths commands
\usepackage{amssymb}	% Extra maths symbols
\usepackage{subfigure}

\def\r500c{R_{\rm 500c}}
\def\R200m{R_{\rm 200m}}
\def\M500c{M_{\rm 500c}}
\def\hiMsun{h^{-1}M_\odot}
\def\Tx{T_X}
\newcommand{\TxM}{T_{X}-M}
\def\Omegasim{{$\textsc{Omega500}$\ }}
\def\chimp{h^{-1}\rm Mpc}
\def\ai{a_{i}}
\newcommand{\Gam}[1]{\Gamma_{\rm 200m}(\ai=#1)}
\def\sp{\rho_{s}}
\def\fnt{f_{\rm nt}}
\def\Pth{P_{\rm th}}
\def\Prand{P_{\rm rand}}

\title[ICM Shapes \& Scaling Relation]{Imprints of Mass Accretion History on the Shape of the Intracluster Medium and the $\TxM$ Relation}

\author[H. Chen et al.]{Huanqing Chen
,$^{1}$\thanks{E-mail: hqchen@uchicago.edu}
Camille Avestruz$^{2,3}$,
Andrey V. Kravtsov$^{1-3}$,
Erwin  T.\ Lau$^{4-6}$, \newauthor
and Daisuke Nagai$^{4,5}$
\\
$^1${Department of Astronomy \& Astrophysics, The University of Chicago, Chicago, IL 60637 U.S.A.}\\
$^2${Enrico Fermi Institute, The University of Chicago, Chicago, IL 60637 U.S.A.}\\
$^3${Kavli Institute for Cosmological Physics, The University of Chicago, Chicago, IL 60637 U.S.A.}\\
$^4${Department of Physics, Yale University, New Haven, CT 06520, U.S.A.}\\
$^5${Yale Center for Astronomy \& Astrophysics, Yale University, New Haven, CT 06520, U.S.A.} \\
$^6${Department of Physics, University of Miami, Coral Gables, FL 33124, U.S.A.}\\
}

\pubyear{2019}

\begin{document}
\label{firstpage}
\pagerange{\pageref{firstpage}--\pageref{lastpage}}
\maketitle

% Abstract of the paper
\begin{abstract}
We use a statistical sample of galaxy clusters from a large cosmological $N$-body$+$hydrodynamics simulation to examine the relation between morphology, or shape, of the
X-ray emitting intracluster medium (ICM) and the mass accretion history of
the galaxy clusters. We find that the mass accretion rate (MAR) of a cluster is correlated with the ellipticity of the ICM.  The correlation is largely driven by material accreted in the last $\sim 4.5$~Gyr, indicating a characteristic time-scale for relaxation of cluster gas. 
Furthermore, we find that the ellipticity of the outer regions ($R\sim\r500c$) of the ICM is correlated with the overall MAR of clusters, while ellipticity of the inner regions ($\lesssim 0.5\r500c$) is sensitive to recent major mergers with mass ratios of $\geq 1:3$. 
Finally, we examine the impact of variations in cluster mass accretion history on the X-ray observable-mass scaling relations. 
We show that there is a {\it continuous\/} anti-correlation between the residuals in the $\TxM$ relation and cluster MARs, within which merging and relaxed clusters occupy extremes of the distribution rather than form two peaks in a bi-modal distribution, as was often assumed previously. Our results indicate the systematic uncertainties in the X-ray observable-mass relations can be mitigated by using the information encoded in the apparent ICM ellipticity.
\end{abstract}

%Our results indicate that the systematic uncertainties in X-ray observable-mass relations can be mitigated by using the information encoded in the apparent ICM ellipticity. 
% We also find that clusters with rounder shapes tend to be more thermalized and therefore hotter at fixed mass, which is consistent with the anti-correlation between ICM ellipticity and cluster MAR. 

% Select between one and six entries from the list of approved keywords.
% Don't make up new ones.

%%%%%%%%%%%%%%%%%%%%%%%%%%%%%%%%%%%%%%%%%%%%%%%%%%

%%%%%%%%%%%%%%%%% BODY OF PAPER %%%%%%%%%%%%%%%%%%

\section{Introduction}
In the hierarchical structure formation scenario of our universe,
small scale density fluctuations on average have larger initial amplitudes than large scale peaks, and thus undergo gravitational collapse earlier.  Continual gravitational collapse leads to the formation of more massive structures. The largest collapsed systems in the present day are galaxy clusters, with total masses of $\approx 10^{14-15}$
$\hiMsun$ \citep[see, e.g.,][for a review]{kravtsov_borgani12}. 

% Filaments and accreting structures - impact on ICM
The tidal fields of the neighboring massive clusters shepherd matter into 
connective filaments \citep{bond_etal96}.  Matter and collapsed haloes accrete on to clusters predominantly along such filaments. Since galaxy clusters grow by accreting materials in their outskirts from the cosmic web, the rate and mode of accretion impact properties of the X-ray emitting cluster gas and of the underlying dark matter \citep[e.g.,][for a recent review and references therein]{walker_etal19}.
%\citep[e.g.,][for a recent review]{rasia_etal04,diemer14,lau_etal15,battaglia15,more_etal15,avestruz16,mansfield17,walker_etal19}.

% Previous studies
Early hydrodynamic simulations of galaxy clusters illustrated how cosmology impacts the mass accretion history and therefore the shape,
or morphology, of the X-ray emitting ICM \citep{evrard_etal93,kasunandevrard_05}.  In
simulations, mass accretion on to a galaxy cluster is often classified
either through the average rate of change in mass over some period of time \citep{diemer14,lau_etal15} or through the mass ratio of merging systems \citep{planelles09,yu15}.  Hydrodynamical simulations have shown that, on average, faster accreting clusters deviate further from dynamical relaxation with higher non-thermal pressure fractions
\citep{lau_etal09,battaglia12,nelson14b,shi15,avestruz16}, underestimate of a cluster mass based on the hydrostatic assumption \citep[e.g.,][for a recent review]{rasia_etal06,nagai07,biffi16,pratt19}, 
and smaller dark-matter splashback radii
\citep{diemer14,adhikari14,more_etal15,shi16,mansfield17} or accretion shock radii \citep{shi16b}.  However, most of the cluster properties examined in these theoretical studies are difficult to observe directly.

% Shapes - studies
One property that is relatively straightforward to infer from observations is the shape of the ICM.  The gas distribution approximately follows equipotential surfaces, thereby probing the shape of the overall potential \citep{lau11,zemp11}. The shape of potential, in turn, reflects the dynamical state of the system.  Dynamically relaxed clusters tend to have rounder mass distributions.  If the total potential is sensitive to the mass accretion history of a galaxy cluster, the mass accretion history can be probed by observations of the ICM shape \citep[e.g.,][]{donahue_etal16,nurgaliev_etal17} and compared with predictions from self-consistent cosmological simulations of galaxy clusters.

% Classification of relaxedness - summary of status
The distribution of ICM shapes impacts cluster selection for cosmology.  Therefore, our understanding of the driving physical mechanisms behind the shape distribution can inform our priors for cluster-based cosmological constraints. Since the dynamical state of a cluster affects our ability to connect observables with the mass of a galaxy cluster, the uncertainties propagate to inferred
cosmological parameters \citep{rasia_etal13,mantz_etal15,lovisari_etal17}.  There has been extensive work to classify clusters according to their dynamic state, particularly in X-ray observations of the cluster gas.  Since we expect more spherical clusters to be dynamically relaxed with relatively little recent accretion, the shape of the ICM has emerged as one of several indicators of relaxedness. Other relaxation criteria include the lack or presence of bright cool cores \citep{santos_etal08} and shifts in the centroid of isophotes in X-ray maps \citep{mohr_etal93}.

% Disturbedness and merger ratio criteria - need to disentangle role
% of major mergers, also role of overall mass accretion history
Shape-based criteria often implicitly assume that non-spherical ICM shapes are a direct result of a recent major merger with some pre-defined mass ratio \citep[e.g.,][]{mathiesenandevrard01}.  However, the definition of a ``major merger''  varies across the literature, indicating the ambiguity behind this definition. In fact, a significant fraction of accreted mass can occur from ``minor mergers" and relatively smooth accretion from cluster-feeding filaments.  Simulations have demonstrated that smooth accretion with sufficiently high momentum density can penetrate into the virialized regions of clusters, potentially impacting the ICM shape \citep{keres05,zinger16a}. The ambiguity in merger definition motivates us to explore different accretion quantities that may impact the ICM shape, including both the continuous mass accretion rate (MAR) and merger definitions of varying mass ratio thresholds.

% Relevance in Tx-M relations
Dynamical classification of clusters can be used to correct observable-mass relations, such as the bias in the $\TxM$ relation.
%\citep[see][for the application to this idea to the Sunyaev-Zel'dovich effect scaling relation]{yu15}.
%\citep{vikhlinin09}.  
The X-ray temperature is a mass proxy that has relatively low scatter of 10\% \citep[e.g.,][]{kravtsov06,vikhlinin09,mantz10}.
%\citet{mathiesenandevrard01} found that 
Hydrodynamical simulations indicate that, for clusters with the same mass, major merger clusters have systematically lower temperature as compared to non-major merger clusters, leading to a mass estimate that is biased low \citep[e.g.,][]{mathiesenandevrard01,nagai07b}. However, this bias is likely induced by the overall high accretion history, thus not necessarily bi-modal. This motivates us to examine the impact of the overall accretion history on the $\TxM$ relation in addition to the merger ratio.

% Summary of what we are addressing
In this work, we use a statistical sample of simulated galaxy clusters extracted
%mass-limited simulated sample of clusters 
from the \Omegasim non-radiative hydrodynamical cosmological simulation to study the relationship between the shape of the ICM and the MAR of galaxy clusters, as measured {\it both} by time averaged mass accretion and merger ratio.  Finally, we assess how the mass accretion history impacts the $\TxM$ relation.

\section{Methodology}
% General simulation description
In this study we use 80 massive clusters from the non-radiative hydrodynamic cosmological simulation, \Omegasim \citep{nelson14a}.  \Omegasim  has a box size length of $500\,\chimp$, and was performed using the Adaptive Refinement Tree code \citep{Kravtsov99,Rudd08}.  The input cosmology is consistent with cosmological constraints from the 5-year data of WMAP mission:  $\Omega_M=0.27$, $\Omega_b=0.0469$, $h=0.7$ and $\sigma_8=0.82$. The simulation follows the evolution of gas and dark matter particles on an adaptive grid, starting with  a base grid of $512^3$ and refining up to eight additional levels, reaching a maximum resolution of $3.8\,h^{-1}$ kpc.  

% Halo finding
Clusters were identified using a halo finder, which first finds a dark matter density peak on a specified scale and then iteratively searches for the center of mass in spheres that increase until the halo radius encloses 500 times the critical density of the universe at the time of analysis. We will denote such radius as $\r500c$. The iterative procedure prevents the halo from mis-centering on a substructure \citep[see][for additional details on halo finder]{nelson12}.

% Merger tree, merger time, merger ratio
Merger trees were constructed from the halo catalog produced by the halo finder by linking haloes between consecutive time snapshots that have $>$10\% of shared dark matter particles.  We calculate merger times by defining two haloes to have merged when their respective spheres of radius $\r500c$ have intersected.  For each merger, there is an associated merger ratio between the primary more massive halo and the second smaller halo.  We use the maximum merger mass ratio to classify accretion modes within a given range of expansion factor $\ai<a<1.0$, where  $\ai$ ranges between $\ai=0.5$ to $\ai=0.9$.  

\subsection{Classification of Mass Accretion Regimes}\label{sec:methods:accretion}
%----------------------------------------------------

% Definition of classification - why we use results to determine classification (arbitrary definitions otherwise), and what the classification is.
Throughout this paper, we define three regimes of mass accretion on to clusters using the maximum mass ratio of mergers that occurred within the last 4.5~Gyr: mergers with mass ratios larger than 1:3 are classified as major, mergers with mass ratios between 1:3 and 1:6 as moderate, and accretion with no mergers of mass ratios larger than 1:6 are classified as smooth. With such classification, our $z=0$ sample of 80 clusters consists of 36 major merger clusters, 20 moderate merger clusters, and 24 clusters that undergo smooth accretion.

% Describe physical intuition for why major mergers vs. moderate mergers vs. smooth accretion would be different, and mention consistency with other studies.
Admittedly, the boundaries between major and moderate mergers, and between mergers and smooth accretion are somewhat arbitrary and these vary in the literature. Broadly, major mergers should involve two systems of comparable mass, significantly impact cluster morphology and happen relatively infrequently. Our choice of major merger definition is consistent with previous studies \citep{planelles09,yu15}. The boundary between mergers and smooth accretion is somewhat more difficult to motivate. Our choice is a trade-off between frequency and the degree of disturbance a merger causes. Mergers with mass ratios less than 1:6 have only minor impact on the overall dynamical state. 
Finally, our specific choice of the time-scale of 4.5~Gyr is motivated from our result in Section~\ref{sec:results:correlation}, showing that matter accreted in the last 4.5~Gyr has the largest impact on the ICM shape. This time scale is also similar to the relaxation time scale found previously by \citet{nelson14a,nelson14b} in detailed analysis of how dynamical state of clusters changes after major mergers.

% Accretion Rate Measurement
We estimate the MAR $\Gamma_{\rm 200m}(\ai)$ within $\R200m$ (the radius enclosing 200 times the mean density of the universe) following \citet{diemer14},
\begin{eqnarray}\label{eqn:gamma_def}
\Gamma_{\rm 200m}(\ai)&=&\frac{\Delta \log(M_{\rm 200m})}{\Delta \log(a)}\\
&=&\frac{\log(M_{\rm 200m_i})-\log(M_{\rm 200m_0})}{\log(\ai)-\log(a_0)},
\end{eqnarray}
where $a_0=1$ is the current expansion factor at which $M_{\rm 200m_0}$ is measured, and $\ai$ is the expansion factor at which $M_{\rm 200m_i}$ is measured.  

\begin{figure*}
  \includegraphics[width=0.5\linewidth]{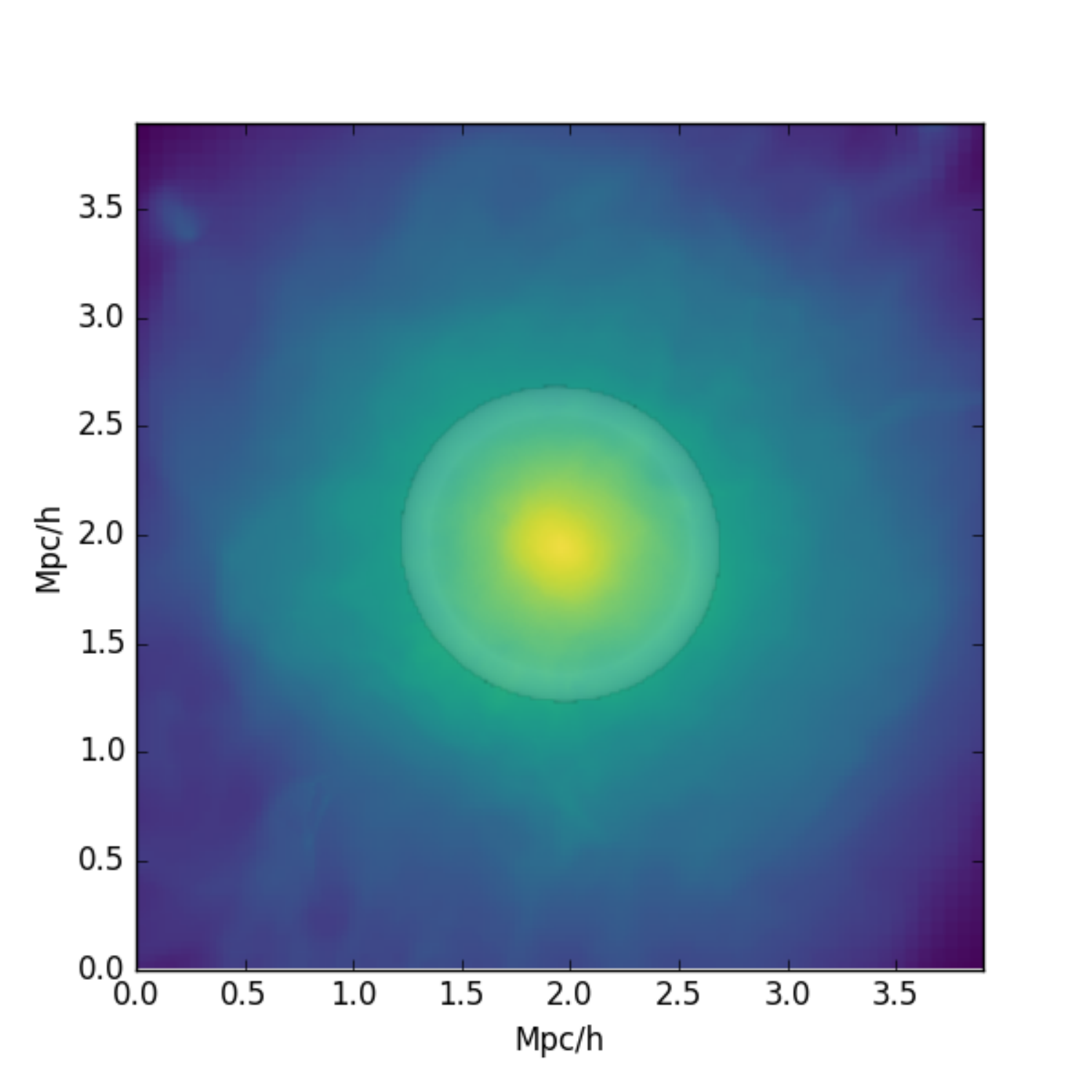}\hspace{-25pt}
  \includegraphics[width=0.5\linewidth]{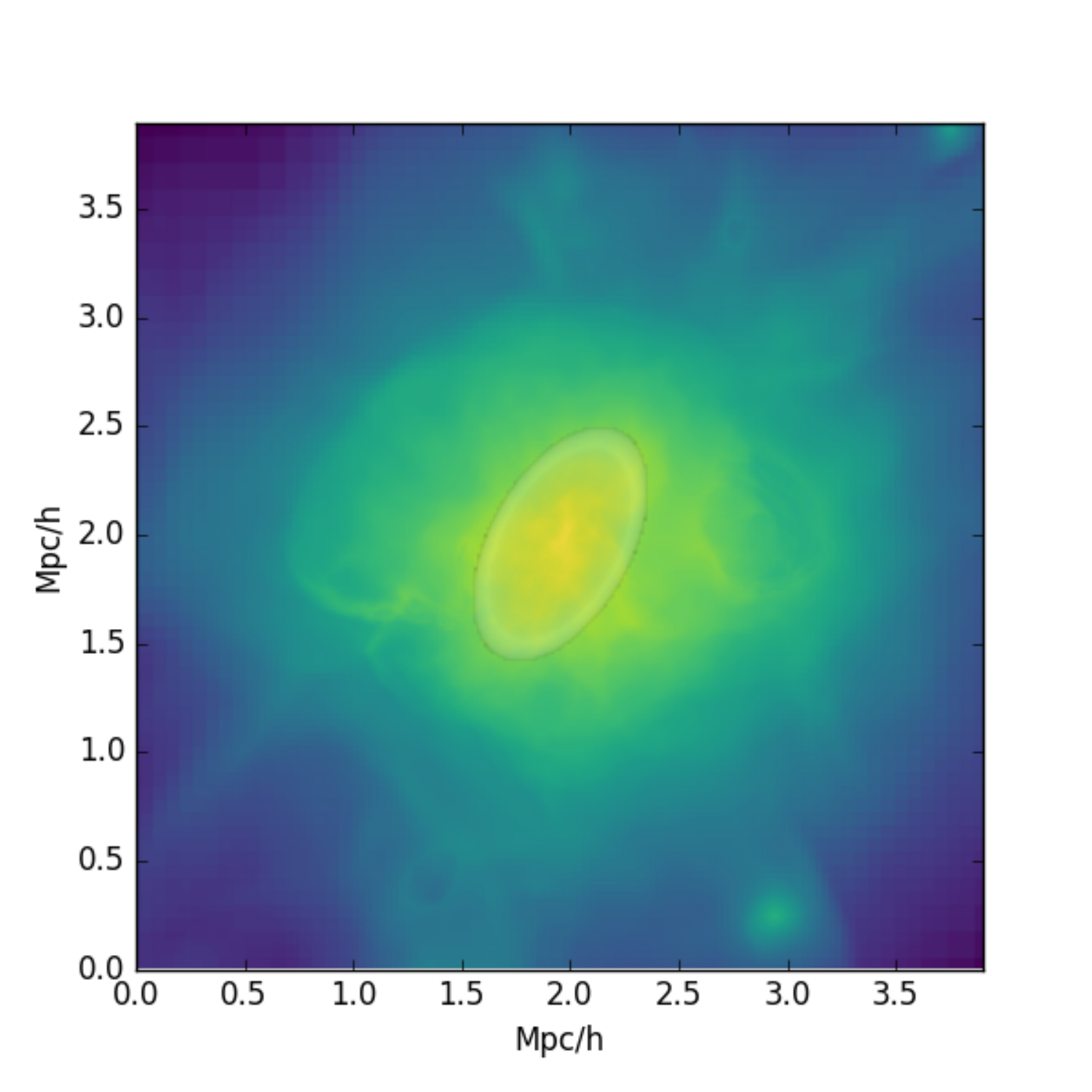}
\caption{Left: Gas density projection  in a region $3.9\,\chimp$ across for a $\M500c=4.3\times10^{14} \hiMsun$ cluster with one of the slowest MARs. The cluster ICM has a measured 3D shape of $c/a=0.9$ at $0.8\r500c$ and accretion rate of $\Gam{0.7}=0.79$. We also overlie the best-fitting ellipsoid at $0.8\r500c$ to show that our best-fitting ellipsoid does describe the ICM shape. Right: Same as the left panel for a $\M500c=2.98\times10^{14} \hiMsun$ cluster that has only undergone moderate mergers since $a_i=0.7$, and has a measured 3D shape of $c/a=0.45$ at $0.8\r500c$ and accretion rate of $\Gam{0.7}=3.3$.}\label{fig:shape_visualization}
\end{figure*}
\subsection{Substructure Removal}\label{sec:methods:substructure}
%---------------------------------------------------------------
Since small gas clumps are prevalent at $R\gtrsim \r500c$, the inclusion of these objects can change the average mass distribution by a large amount and significantly bias the measured ellipticity. We therefore remove the substructure using the method described in \citet{zhuravleva13}. 
We use a value for $f_{\rm cut}=3.5$, which removes most substructure components while leaving bulk components of the main cluster unaffected. To calculate the ellipticity, we replace the removed substructure with gas whose density matches the median gas density at that cluster-centric radius.  When calculating the moment of inertia, we use the halo center as the fixed cluster center.

\subsection{Ellipticity Measurement}\label{sec:methods:ellipticity}
%------------------------------------------------------------------

We estimate the ellipticity of the ICM in each cluster using the moment of inertia tensor method \citep[e.g.,][see \citealt{zemp11} for a detailed techincal discussion of how such estimate should be done in practice]{dubinski_carlberg91}.  We first calculate the shape tensor, $S_{ij}$, in a spherical shell,
$$ S_{ij}={\frac{\Sigma_k m_k(\boldsymbol{r}_k)_i (\boldsymbol{r}_k)_j }{\Sigma_k m_k}} 
$$
where the $i$th and $j$th indices correspond to the $x$, $y$, or $z$ axes of the shape tensor and $k$ denotes the index of the gas cell.  If the mass distribution of an object is close to an ellipsoid, then the eigenvalues and eigenvectors of this tensor provide the estimated axis ratio and orientation of the ellipsoid.  Next, we calculate the tensor in the estimated ellipsoidal shell and iterate over this process until the axis ratios converge.  We define the cluster shape at each radial bin with respect to $\r500c$, with the cluster shape defined as the ratio between the semi-minor axis and the semi-major axis, $c/a$.

\begin{figure*}
\begin{center}
\includegraphics[width=0.45\linewidth]{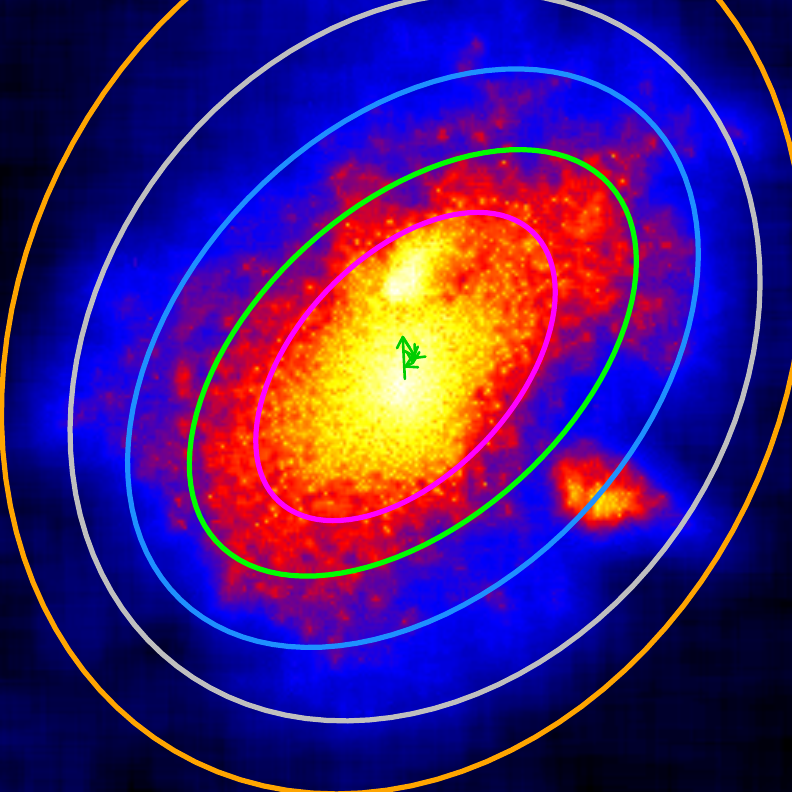}\hspace{1pt}%scale=0.33
\includegraphics[width=0.45\linewidth]{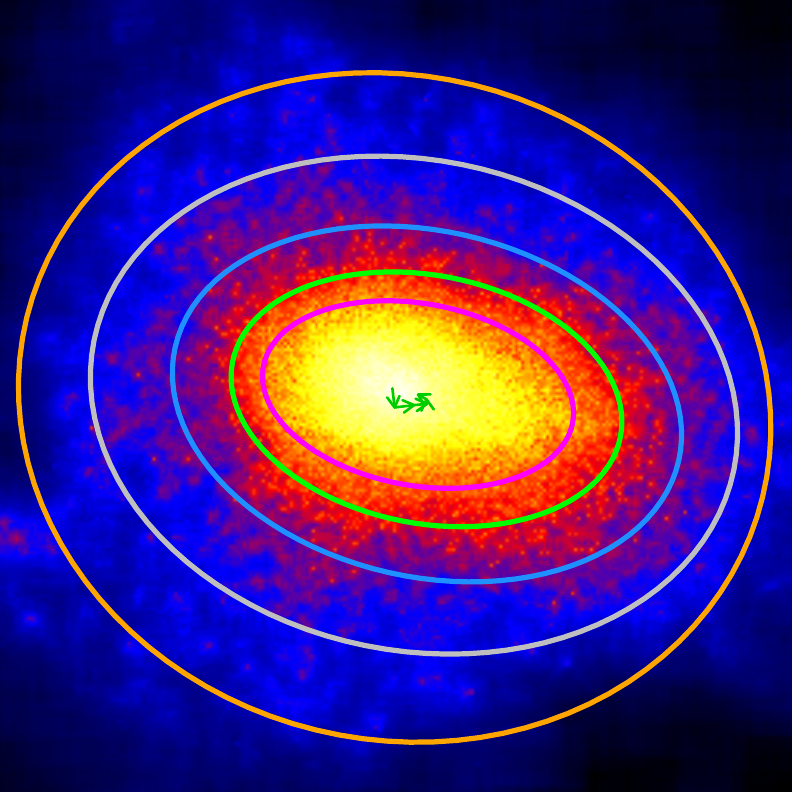}\hspace{1pt}%scale=0.45
%scale=0.33
\vspace{2pt}
\caption{Mock {\em Chandra} X-ray maps of clusters with varying accretion histories. Each ellipse is best-fitted to the image X-ray isophotes using the SPA code. Green arrows in the center show the shift between the X-ray peak and center of the best-fitting ellipse in each image. Left: CL82, a fast but smoothly accreting cluster, with only two minor mergers of mass ratio of 1:10 in the last 4.5~Gyr. The most recent major merger event happened 9~Gyr in its past. The green ellipse corresponds to $0.8\r500c$. Right: CL34, a major merger cluster. This cluster experienced a major merger of $1:1.6$ at $a=0.8$. The gray ellipse corresponds to $0.8\r500c$. The similarity of the appearance of these two clusters shows that fast accretion can driven elongation of the ICM like major merger. However, examination of the cores reveals that the major merger cluster has more elongated core compared with the rounder and more intact core of the smoothly accreting cluster. }
\label{fig:CL82_isophote}
\end{center}
\end{figure*}
\begin{table*}
\centering
\caption{Characteristics of the mock X-ray clusters in Figure~\ref{fig:CL82_isophote} }
\begin{tabular}{ccccccc}
\hline
\hline
 & $\M500c[\hiMsun]$ & $\r500c[h^{-1}{\rm kpc}]$ 
 & 3D ellipticity  & $\Gam{0.7}$ 
 & Max merger ratio & SPA ellipticity   \\
 &   & 
 & at $0.8 \r500c$ & 
 & between ($0.7<a<1$)& at $R/R_{500c}=0.8$\\
\hline
CL82 & $1.8\times 10^{14} $&679  & 0.54 & 2.3 & 1:10 & 0.60 \\
CL34 & $3.8\times 10^{14} $&870  &0.54 & 2.8& 1:1.6 & 0.67 \\[0.5mm]
\hline
\end{tabular}
\label{tab:spa_clusters}
\end{table*}

% Sample figure description
Figure~\ref{fig:shape_visualization} shows the gas density projection of two example clusters and the corresponding projected ellipse from the best-fitting ellipticity measurement made at $0.8 \r500c$.  The left panel shows a cluster of mass
$\M500c=4.3\times10^{14}\hiMsun$ with one of the slowest MARs ($\Gam{0.7}=0.79$) among systems in our sample.  Due to low overall accretion and merger rates the ICM has a round shape with ellipticity measurement of $c/a=0.9$ at $0.8 \r500c$.  The right panel shows a cluster accreting mass with a faster rate of $\Gam{0.7}=3.3$ that has not undergone any major mergers since $\ai=0.7$.  The right hand cluster has a mass of $\M500c=2.98\times10^{14} \hiMsun$, and an ellipticity measurement of $c/a=0.45$ at $0.8 \r500c$.  Note, the elliptical shape of the cluster on the right-hand-side is therefore not caused by recent major mergers.

\subsection{Measuring Integrated ICM Observable $\Tx$}

% Examine the relationship between accretion rate, mode of accretion, shape, and scatter in X-ray observable.  
Both the accretion rate and mode of accretion could potentially drive the scatter in the observable-mass relation of galaxy clusters.  Selecting clusters by their shapes may help us reduce this scatter. This motivates our study of the integrated ICM observable $\Tx$ in order to investigate the relationship between scatter and accretion. 

The spectroscopic-like temperature has been shown to be a more accurate estimation of the observed X-ray temperature of the ICM \citep{mazzotta04,vikhlinin06}.  We therefore use the definition of spectroscopic-like weighting for $\Tx$, 
$$\Tx=\frac{\sum_{i} T_i^{0.25} \rho_i^2 \Delta V_i}{\sum_{i} T_i^{-0.75} \rho_i^2 \Delta V_i} $$
where $T_i$, $\rho_i$, $V_i$ are the temperature, density and volume of the cell, and the sum is over a sphere from the center to $\r500c$.  
We exclude the core region up to $0.15\r500c$, as is customary in both theoretical and observational analyses  \citep[e.g.,][]{kravtsov06,vikhlinin09,fabjan11,planelles14}. We have also tested other definitions of $\Tx$, including those weighted by mass and emissivity, with and without core, and our main conclusions remain robust.

%%%%%%%%%%%%%%%%%%%%%%%%%%%%%%%%%%%%%  
\section{Results}
%%%%%%%%%%%%%%%%%%%%%%%%%%%%%%%%%%%%% 

%------------------------------------------------
\subsection{Merger-driven vs Smooth Mass Accretion}\label{sec:results:mergervssmooth}
%------------------------------------------------
% Summary of subsection 
Results in this section will be presented for the entire cluster sample, as well as for subsamples that exclude major and major$+$moderate mergers. Some of the clusters in the latter ``smoothly accreting'' sample have MARs on par with the clusters that underwent recent major mergers. As we embark on the exploration of the correlations of the ICM shape with MAR, it is worth examining the effects of mergers versus smooth accretion in some details for individual clusters.

Figure \ref{fig:CL82_isophote} shows mock {\em Chandra} X-ray images of two clusters from our sample with properties summarized in Table~\ref{tab:spa_clusters}. The mock X-ray maps were produced using the {\em SOXS} package.\footnote{\url{http://hea-www.cfa.harvard.edu/~jzuhone/soxs/}} We use the SPA code\footnote{SPA stands for Symmetry, Peakiness and Alignment. SPA code website: \url{https://sites.google.com/site/adambmantz/work/morph14/}}, described in \citet{mantz_etal15}, to fit the isophotes for five different surface brightness values.  The best-fitting isophotes are shown as the overlaid ellipses of different colours. Green arrows in the center of the images show the shift between the X-ray peak and centers of the ellipses.

Both clusters were selected to have similarly high MARs of $\Gam{0.7}=2.8$ and $\Gam{0.7}=2.3$.  The former, CL34, accreted most of the mass in a major merger. The other, CL82, accreted most of its mass smoothly, only experiencing mergers smaller than 1:10 in the same time interval.

%%%%%%%%%%%%%%%%%%%%%%%%%%%%%%%%%%%%%%%%%%%%%%%%%
\begin{figure}
\begin{center}
\includegraphics[width=\columnwidth,trim={0 0 0 0},clip]{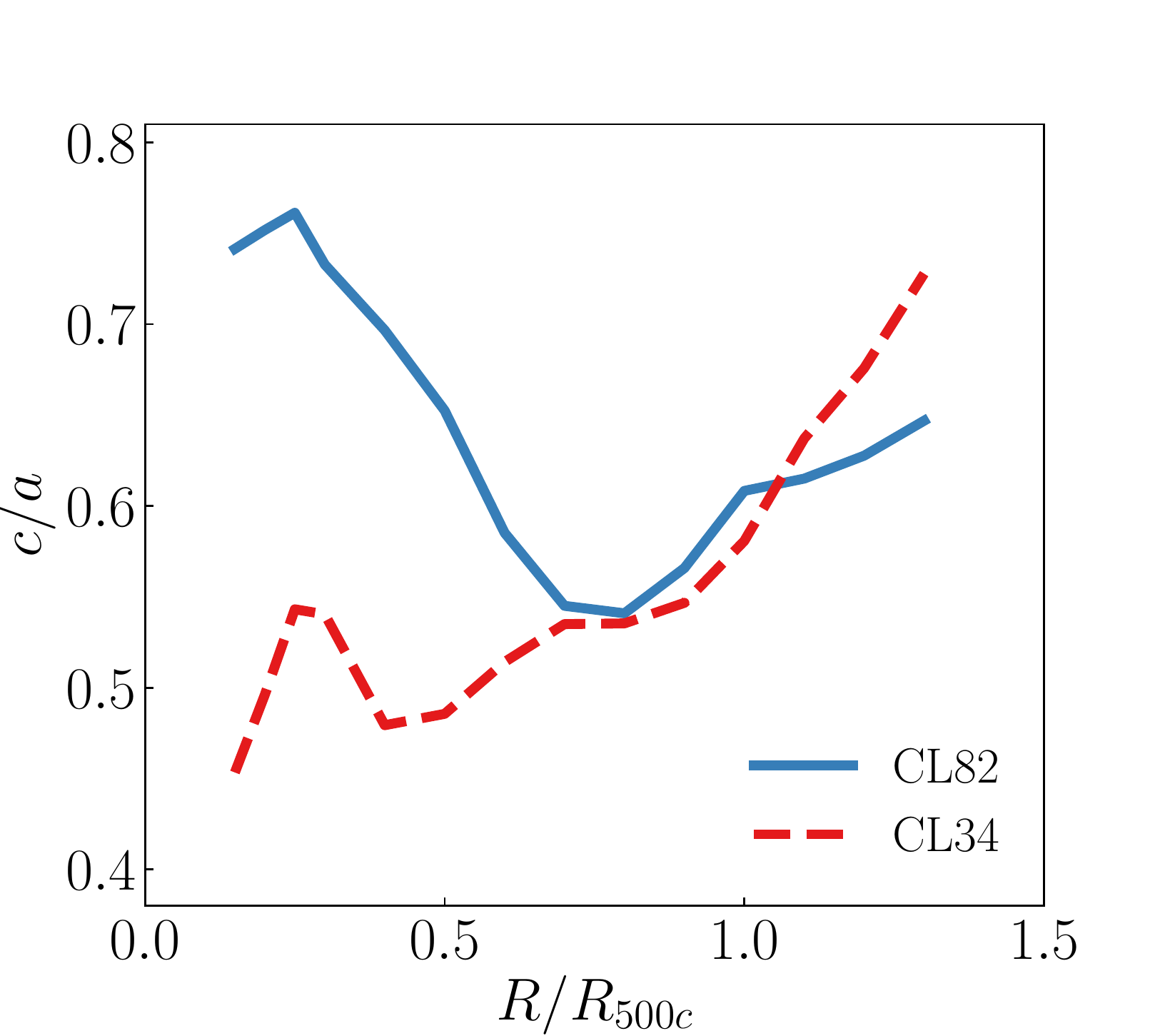}
\caption{3D radial profiles of axis ratios $c/a$  of the two clusters shown in Fig.~\ref{fig:CL82_isophote}. The figure shows that the cluster with a recent major merger (dashed red line) has a more elliptical core. In contrast, the two clusters have comparable ICM ellipticity at outer radii $\sim \r500c$.}
\label{fig:twoCL_coa_prof}
\end{center}
\end{figure}
%%%%%%%%%%%%%%%%%%%%%%%%%%%%%%%%%%%%%%%%%%%%%%%%%

% First cluster
Figure~\ref{fig:twoCL_coa_prof} shows the radial shape profile of both clusters for comparison.  Although the cluster CL82 has not experienced any major merger within recent 9~Gyr, its ICM shape is quite elliptical: the 3D axis ratio of the best-fitting ellipsoid is $c/a=0.54$ at $0.8\r500c$ (see blue solid line in Figure~\ref{fig:twoCL_coa_prof} for the full radial profile of the measured shape).  Its X-ray isophote at that same radius has an ellipticity of $0.60$, illustrated in the right panel of Figure~\ref{fig:CL82_isophote}.  Note that X-ray isophotes measured by the SPA code are rounder than the 3D measurements due to projection effects. Finally, the X-ray image of CL82 also visually appears to be disturbed and unrelaxed as well as elongated despite a lack of a recent major merger.

% Second cluster
For comparison, CL34 has experienced a 1:1.6 major merger at $a=0.8$ (i.e., $\approx 2.96$ Gyr ago).  It has a 3D ellipticity of $c/a=0.54$ when measured at $0.8\r500c$ (see red dashed line in Figure~\ref{fig:twoCL_coa_prof} for full radial profile of the measured shape), which is similar to that of the fast smoothly accreting cluster, CL82. The X-ray isophote at $0.8\r500c$ in this projection has an ellipticity of $0.67$.  

The case example of CL82 shows that high MARs are not necessarily associated with major mergers. Moreover, the visual ellipticity and apparent relaxation state of the ICM in X-ray images may be comparable for a smooth fast-accreting cluster, such as CL82, as for a post-major merger cluster, such as CL34.  

% Difference between the two
One striking difference between the two clusters is in the ellipticity of the ICM at the innermost regions. The 3D $c/a$ radial profiles of the clusters in Figure~\ref{fig:twoCL_coa_prof} shows that CL34 has a more elongated inner shape than CL82 at $R/\r500c\lesssim 0.8$.  This illustrates how the major merger substantially impacted the inner ICM region.  On the other hand, the overall MAR appears to drive the ellipticity of the intermediate and outer regions, regardless of the mode of accretion.  Indeed, as we will see in the next section, there is a {\it continuous trend} that relates the ICM shape at large radii with the MAR that is independent of accretion mode. 

%-----------------------------------------------------------
\subsection{Correlation between the MAR and ICM Ellipticity}\label{sec:results:correlation}
%-----------------------------------------------------------

\begin{figure*}
 \includegraphics[width=0.5\linewidth]{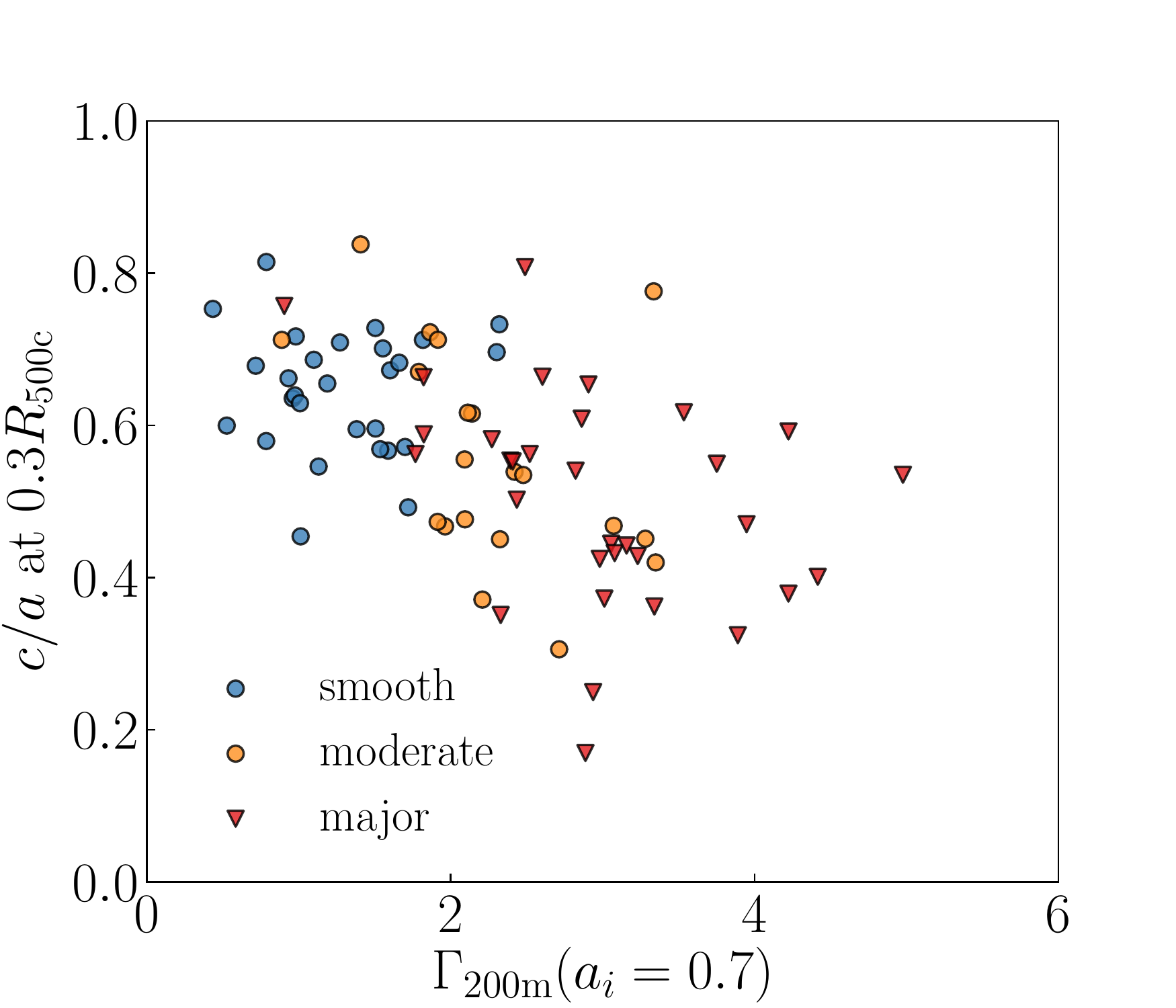}\hspace{-10pt}
  \includegraphics[width=0.5\linewidth]{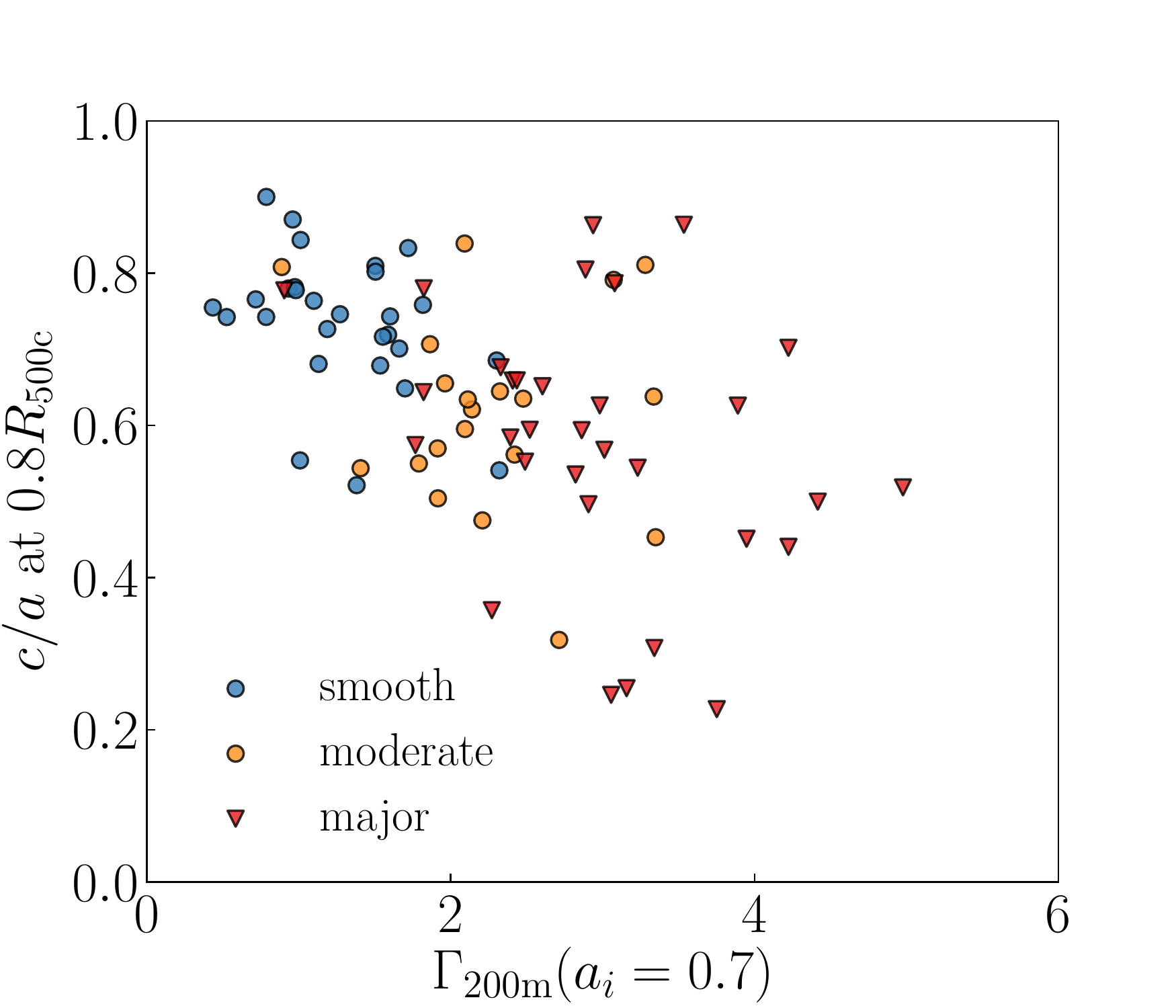}\hspace{-10pt}

\caption{Left: 3D axis ratio measured at the inner radius of $0.3\r500c$ vs. MAR, $\Gam{0.7}$. Right: the same, but for the
axis ratio measured at $0.8\r500c$. The Spearman coefficient for these two quantities is $-0.59^{+0.08}_{-0.06}$ and  $-0.51^{+0.12}_{-0.08}$ for $0.3\r500c$ and $0.8\r500c$, respectively. The trend in the right panel is {\it not} affected by the exclusion of major mergers. See also Figure \ref{fig:spearman_vs_radius_gamma_dep}.}
\label{fig.scatterplot}
\end{figure*}

Figure \ref{fig.scatterplot} shows the anti-correlation between the MAR, $\Gam{0.7}$, and the axis ratio of the best-fitting 3D ellipsoid measured at $0.3\r500c$ (left) and at $0.8\r500c$ (right). 
Although the scatter in this relationship is substantial, there is a clear anti-correlation 
between the MAR and ICM ellipticity: the Spearman's rank coefficients are $\sp=-0.59^{+0.08}_{-0.06}$ and $-0.51^{+0.12}_{-0.08}$ (with errors calculated from bootstrapping) for ellipticities measured at  $0.3\r500c$ and $0.8\r500c$, respectively. When we exclude major mergers, corresponding to the red triangles, the correlation slightly weakens to $\sp=-0.44^{+0.09}_{-0.17}$ at $0.3 \r500c$, but the decrease is not statistically significant. On the other hand, the shape of the ICM at outer radius $0.8 \r500c$ is sensitive to the overall rate of mass accretion, not to the mode of accretion. After removing major merger clusters in the sample the Spearman coefficient for the shape measured at $0.8 \r500c$ does not change significantly (from $-0.51^{+0.12}_{-0.08}$ to $-0.49^{+0.13}_{-0.12}$).  If we further examine the subsample of clusters with only smooth accretion (blue circles), the correlation at $0.8 \r500c$ remains ($\sp=-0.48^{+0.27}_{-0.06}$).

%%%%%%%%%%%%%%%%%%%%%%%%%%%%%%%%%%%

%%%%%%%%%%%%%%%%%%%%%%%%%%%%%%%%%%%
\begin{figure*}
\includegraphics[width=1\linewidth]{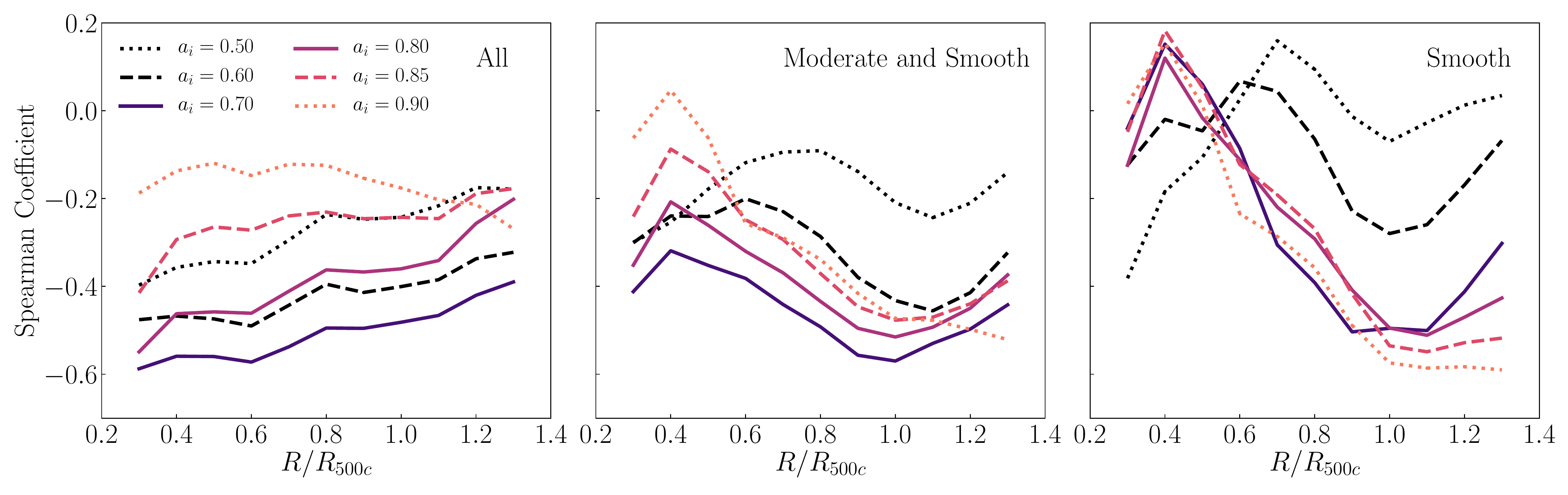}\hspace{-10pt}

\caption{The Spearman's rank correlation coefficient between the cluster accretion rate, $\Gamma_{\rm 200m}(\ai)$, and the cluster ellipticity, indicated by the axis ratio $c/a$ for cluster samples with different accretion modes (from left to right: the entire sample, clusters with moderate and smooth accretion, clusters with smooth accretion).   Each line colour corresponds to different values of $\ai$. 
For the entire sample (``all''), the strongest correlation occurs when we measure the axis ratios at $0.3<R/\r500c<1.0$ and with an initial expansion factor of $\ai=0.7$, indicating a characteristic time-scale for ICM shape relaxation.  The anti-correlation remains strong between $0.8\lesssim R/\r500c\lesssim1.0$, indicating the overall MAR, not mergers, is responsible for the shape at outer radii.  The characteristic time-scale for accretion of $a_i=0.7$ remains the same.  It is worth noting that the bin-to-bin fluctuations of Spearman coefficients are smaller than the bootstrap errors since the shape measurements in neighboring radial bins are correlated.  The radial bin correlation propagates to the correlation in Spearman coefficients, but shape measurements at radii separated by $\gtrsim 0.5 \r500c$ are not correlated. 
}\label{fig:spearman_vs_radius_gamma_dep}
\end{figure*}
%%%%%%%%%%%%%%%%%%%%%%%%%%%%%%%%%%%

% Characteristic timescale for shape relaxation: Cluster shape remembers matter accreted within a certain timescale
Given that clusters with more recent accretion and merger events have  more elliptical shapes, the anti-correlation between ellipticity and the MAR should depend on the time range over which the MAR is measured.  To this end, we explore how the correlation strength varies with the time interval range used to define the accretion rate.  We also explore the strength of the correlation at different cluster-centric radii to find the radii where the ICM shape correlation with the MAR is the strongest for each time interval. 

We measure mean cluster MARs, $\Gamma_{\rm 200m}$, using the time interval from an initial expansion factor ranging between $0.5\leq a_i\leq 0.9$ and the current epoch ($a=1$).  The time difference between expansion factors of $a= 0.5-1$ corresponds to $\approx 7.84$ Gyr.  The range of time intervals used to define MARs also includes the $\sim 4$ Gyr relaxation time-scale for gas motions after a 1:1 major merger \citep{nelson14b}. We measure the 3D ellipticity of the ICM at radii ranging between $0.3\leq R/\r500c\leq 1.3$. The ICM shape at these radii is relatively smooth and can be described reasonably well by an ellipsoid. For each choice of the time interval for MAR and radius for shape measurement, we calculate the Spearman's rank correlation coefficient between the cluster shape and MAR. 

% First figure
Figure~\ref{fig:spearman_vs_radius_gamma_dep} shows the Spearman's rank correlation coefficient as a function of the radius at which the ICM ellipticity is measured.  Each line corresponds to a different value of $a_i$ in measuring the MAR. The left panel shows this relationship for all clusters in the sample, while the middle and right panels respectively show the relationship in the absence of major mergers and in the absence of major$+$moderate mergers.  We briefly note that the bin-to-bin fluctuations in Figure \ref{fig:spearman_vs_radius_gamma_dep} is smaller than the typical bootstrap errors.  Since  the shape measurement of neighboring bins are highly correlated, the corresponding Spearman coefficients in neighboring bins are also correlated, leading to smoother fluctuations between radial bins.

% Timescale findings
For the entire sample (left panel), the correlation is strongest for $a_i=0.7$ at all measured radii, which means that mass accretion in the interval of $0.7<a<1.0$ leaves the strongest imprint on the ICM shape.  The lookback time corresponding to this interval is $\approx 4.54$ Gyrs, which is close to post-merger dynamical relaxation time of gas motions predicted by simulations \citep{nelson12,nelson14b,zhang16}.  The ICM shape does not retain as much information from accretion and mergers on longer timescales, e.g. prior to $a_i=0.7$, due to relaxation.  

On the other hand, accretion defined for a shorter time interval does not include all the information that impacts the ICM shape. 
Accretion that occurs within $\leq2$~Gyr (time between $\ai=0.9$ and $a_0=1.0$) does not have enough time to affect the morphology of most cluster centers, but impacts shape around $\r500c$.

In the middle panel of Figure~\ref{fig:spearman_vs_radius_gamma_dep}, we exclude major mergers in the sample. Again, it shows that the correlation between shape and accretion rate is strongest for accretion measured since $a_i=0.7$, but the radial dependence changes. Here, we see that the anti-correlation between accretion rate and shape weakens at inner radii, but at $\approx \r500c$, gas shape is still strongly anti-correlated with the MAR. 
As we additionally remove moderate mergers (right panel), the correlation between $\Gam{0.7}$ and shape at inner radii further weakens while at outer radii still remains strong.
This is likely because smooth accretion contributes a larger fraction of mass at the outer radii and thus affects the ICM shape there more significantly \citep[e.g.,][]{tormen04}. 

%AK I think this is extraneous in the conclusions. This should be discussed in discussion.
%
%-------------------------------------------------
\subsection{The Effects of Accretion and Shape Criterion on $\TxM$ relation}\label{sec:results:TxM}
%---------------------------------------------------
% Motivation for relaxedness criteria 

One of the most interesting aspects of the ICM ellipticity--accretion rate correlation explored above is that the ICM shape is not just sensitive to recent mergers, but exhibits a {\it continuous correlation with the MAR}.  Given that deviations from a spherical ICM distribution is one of the sources of scatter in the observable-mass relations of clusters, we now explore how the scatter in the $\TxM$  relation is affected by the accretion rate and corresponding ellipticity of the cluster ICM.  We focus on the $\TxM$ relation here because the  scatter of this relation is particularly sensitive to the cluster dynamical state \citep{mathiesenandevrard01}.

The $\TxM$ relation of clusters can be well described by a power-law with log-normal intrinsic scatter \citep[e.g.,][]{evrard14}. We fit a power-law model with intrinsic scatter to the values of $\log_{10}(\Tx [{\rm K}]) $ and $\log_{10}(\M500c [\hiMsun])$ for clusters in our sample.  We use a Markov chain Monte Carlo (MCMC) sampler, implementing the affine-invariant algorithm of \citep{GW10} with free parameters -- slope, $m$, intercept, $c$, and intrinsic Gaussian scatter, $\sigma$:
$$ {\mathcal{L}}(d|m,c,\sigma)=\prod_{i}\, \frac{1}{\sqrt{2\pi\sigma^2}}\,\exp{\left[-\frac{(y_i-mx_i-c)^2}{2\sigma^2}\right]},$$ 
We adopt bounded flat priors on the angle of the line with respect to x-axis, distance from original point to the line, and intrinsic scatter \citep[see, e.g.,][]{sharma17}. 

To remove the degeneracy between slope $m$ and intercept $c$, we choose a pivot point to be close to the median of the data points. Therefore, $y_i$ here is $\log_{10}(\Tx [{\rm K}]) - 7.7 $, $x_i$ is  $\log_{10}(\M500c [ \hiMsun]) - 14.6$, and the model function is:
\begin{eqnarray}
\log_{10}(\Tx [{\rm K}]) - 7.7 = m (\log_{10}(\M500c [\hiMsun])-14.6) - c
\end{eqnarray}
The best-fitting parameters are $m=0.61\pm0.05$, $c=-0.009\pm0.009$, and $\sigma=0.075\pm0.006$. The slope is consistent with the self-similar value of $2/3$, as could be expected, given that our simulation does not include radiative cooling.

\begin{figure*}
    \centering
    \mbox{\subfigure{\includegraphics[width=3in]{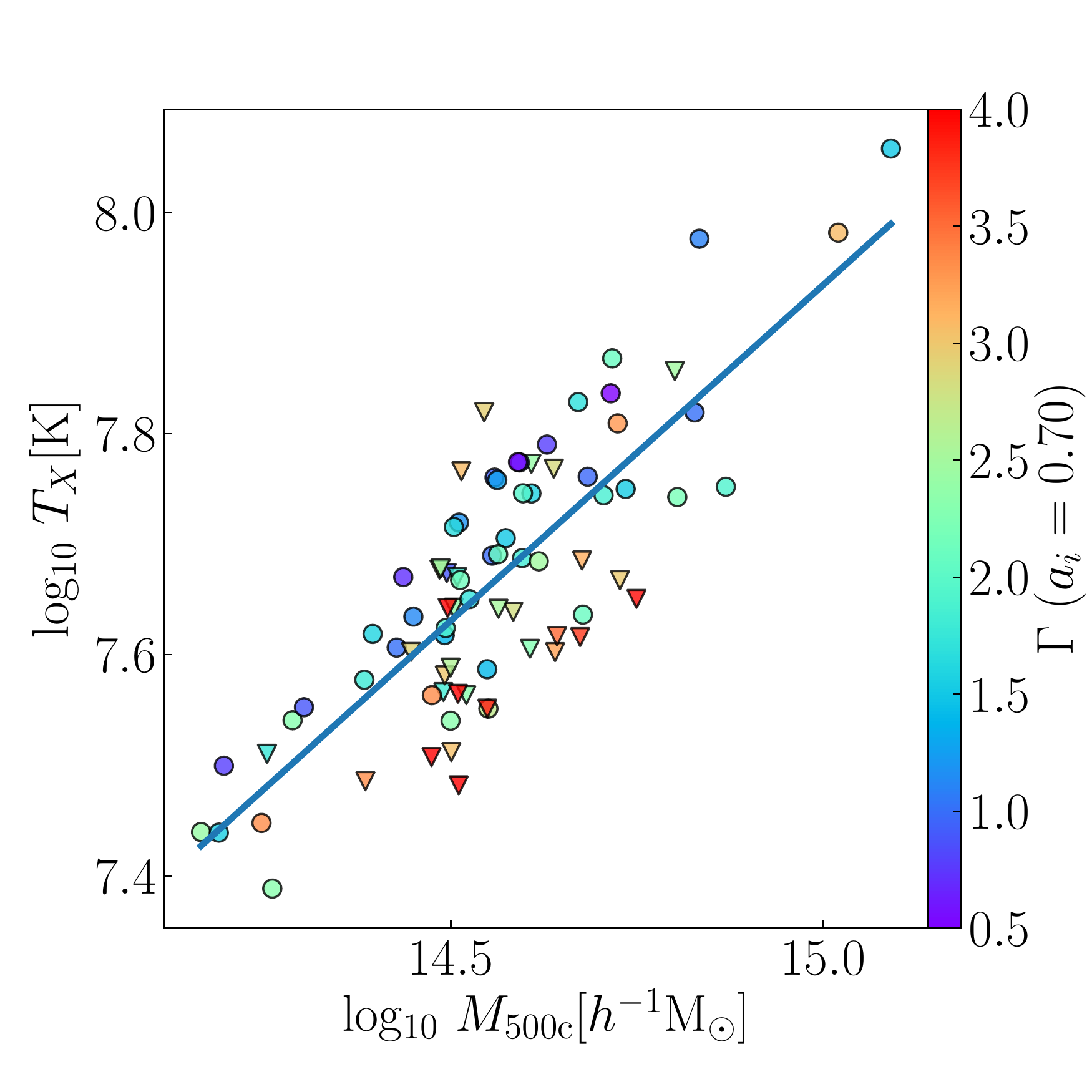}}\quad
\subfigure{\includegraphics[width=3in]{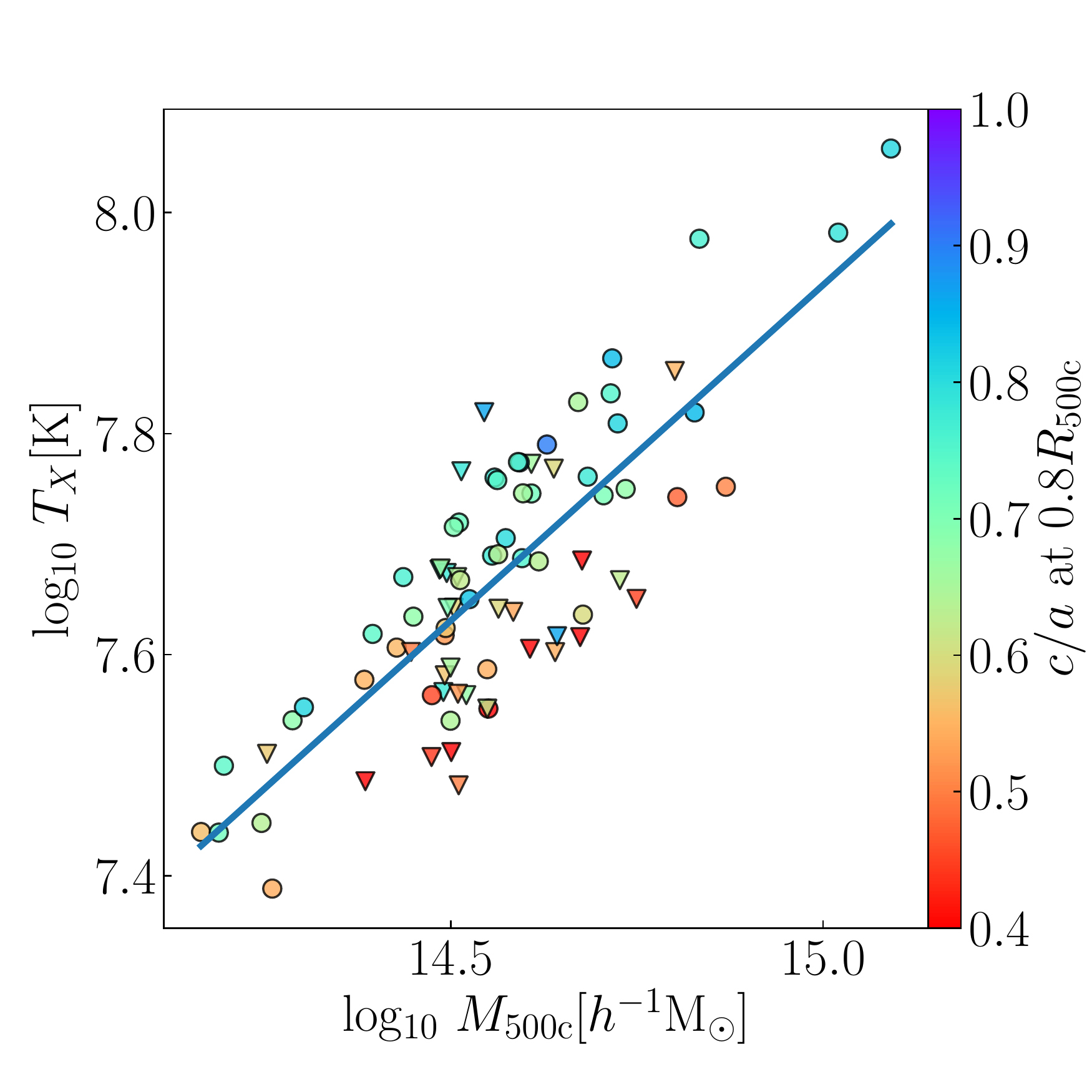}}}\\\vspace{-10pt}
    \mbox{\subfigure{\includegraphics[width=3in]{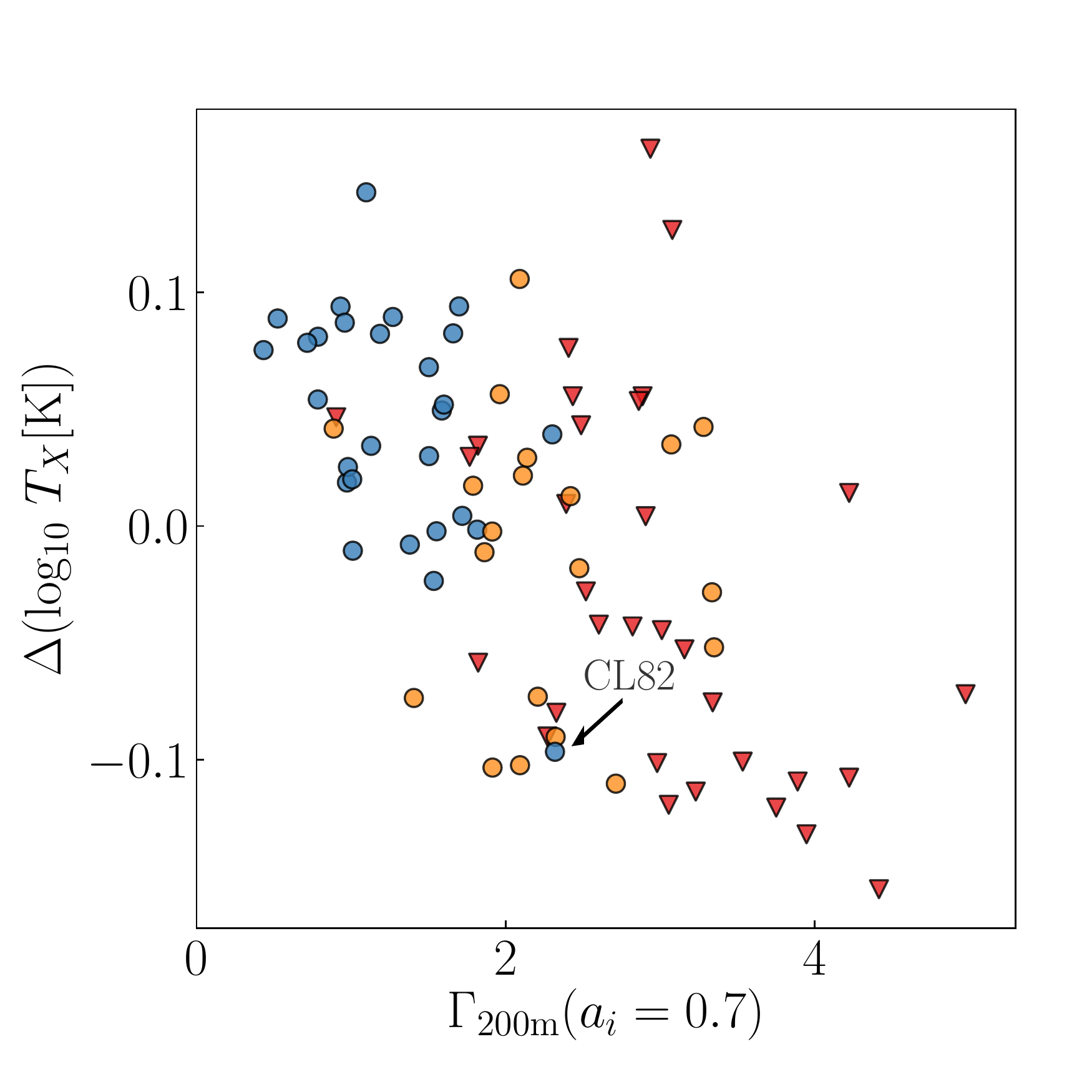}}\quad
	\subfigure{\includegraphics[width=3in]{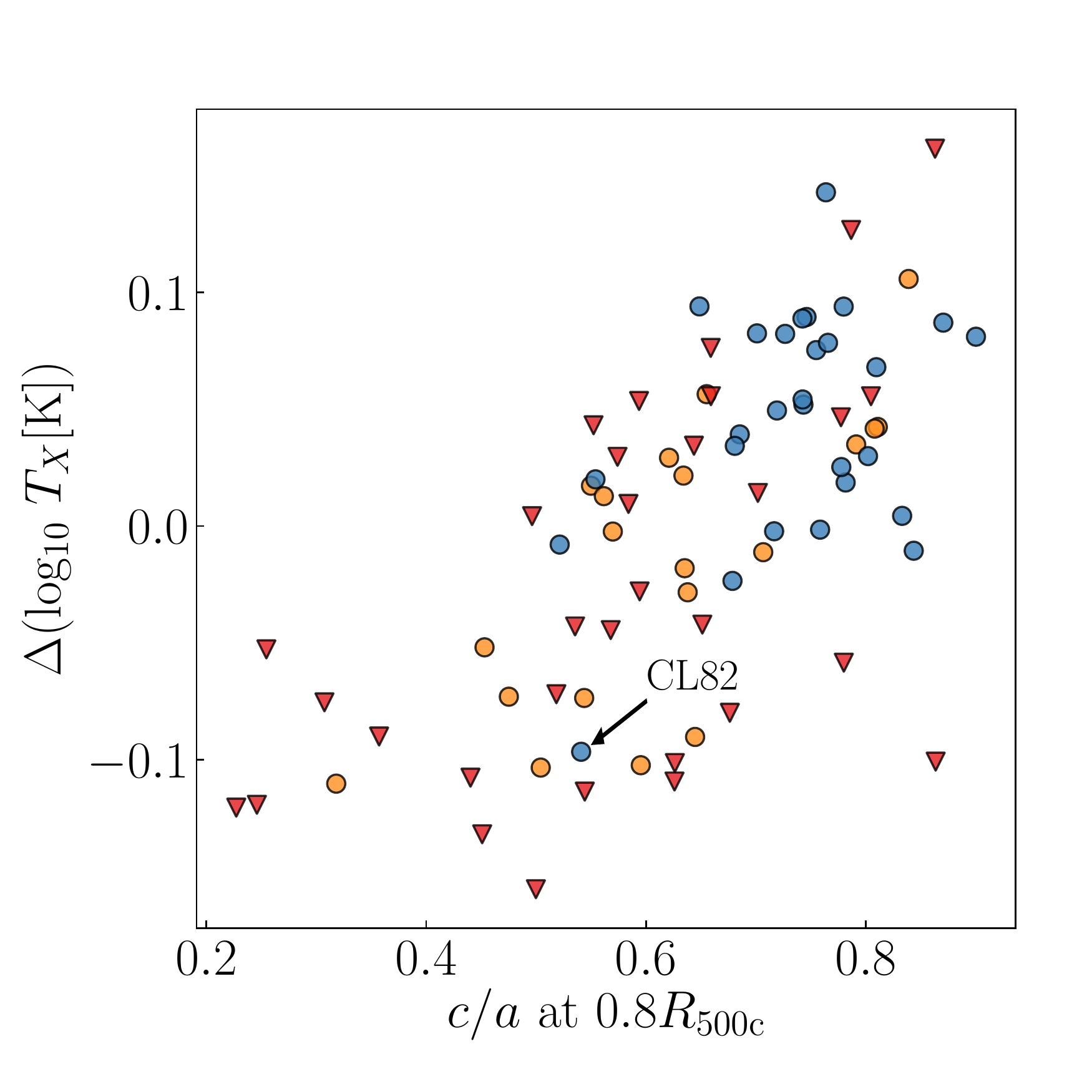}}}
        \caption{Upper panels: Scaling relation between $\Tx$ and $\M500c$, colour-coded by $\Gam{0.7}$  (left) and 3D ICM shape $c/a$  measured at $0.8\r500c$ (right). The blue line corresponds to the best-fitting linear relation. We can see that clusters with lower accretion rates and larger $c/a$ ratios mostly sit above the best-fitting line. Inversely, clusters with higher accretion rates and smaller $c/a$ ratios sit below the best-fitting line. Lower panels: Residual from the best-fitting values in the upper panels, colour-coded by accretion modes (labels are the same as in Figure~\ref{fig.scatterplot}: red triangles -- major merger clusters, orange circles -- moderate merger clusters, blue circles -- smooth accreting clusters). The bottom left figure shows a smooth trend of temperature residuals with MAR, regardless of accretion mode. The bottom-right panel shows  a similarly continuous trend of residuals with cluster ellipticity.  
}\label{fig:txscaling}
\vspace{10pt}
\end{figure*}

The upper panels of Figure~\ref{fig:txscaling} show the scaling relation between the 3D spectroscopic-like, core-excised temperature and  $\M500c$. We colour-code the data points in the upper left panel by the accretion rate, $\Gam{0.7}$.
The lower left panel of Figure~\ref{fig:txscaling} shows the corresponding residuals of individual clusters from the best-fitting power law relation as a function of $\Gam{0.7}$, colour-coded by the accretion mode defined by merger classification as shown in the legend of Figure~\ref{fig.scatterplot}.  The residuals of the $\TxM$ relation decrease systematically as $\Gam{0.7}$ increases.  Although previous studies have found similar correlations of residuals with the merging state of the cluster  \citep[e.g.,][]{mathiesenandevrard01,kravtsov06,nagai07b}, here we show that 
merging clusters do not constitute a separate population distinct from smoothly accreting clusters.  Instead, there is a continuous trend between the residual of the $\TxM$ relationship and the MAR, regardless of accretion mode \citep[see also][who showed that high $\Gamma_{\rm 200m}$ clusters have less thermalized gas]{avestruz16}. 

For example, there are clusters with moderate (orange circles) and major (red triangles) mergers that have slow accretion rates ($\Gam{0.7}\leq2$) and a positive residual, meaning their temperatures are higher than the best-fitting relation prediction. There are also clusters with high accretion rate ($\Gam{0.7}\geq2$) that are smoothly accreting mass (blue circles) with a negative residual, meaning their temperatures are lower than predicted by the best-fitting relation.  In particular, CL82 discussed in section \ref{sec:results:mergervssmooth} is the fastest smoothly accreting system. Its residual is negative. 

The negative residuals of clusters with high MAR may be due both to the larger fraction of cool gas in accreted substructures and non-thermalized gas motions.  However, when we use the substructure-excised and core-excised $\Tx$ value, fast accreting clusters still tend to have negative residuals.  This indicates that the primary cause of the lower than average $\Tx$ value in a fast accreting cluster is the large fraction of gas with non-thermalized kinetic energy \citep{avestruz16}.

\begin{figure*}
  \includegraphics[width=0.343\linewidth]{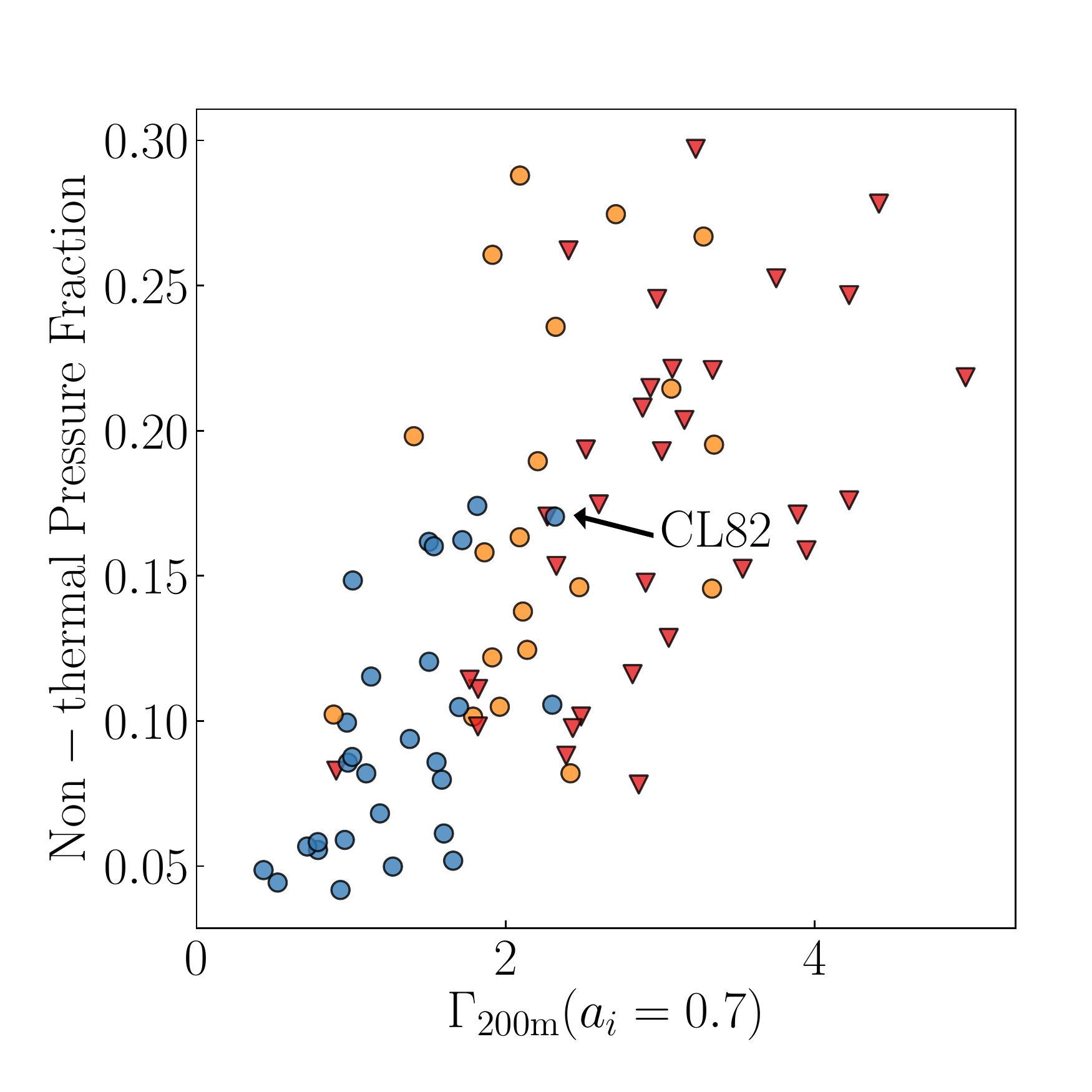}\hspace{-10pt}
  \includegraphics[width=0.343\linewidth]{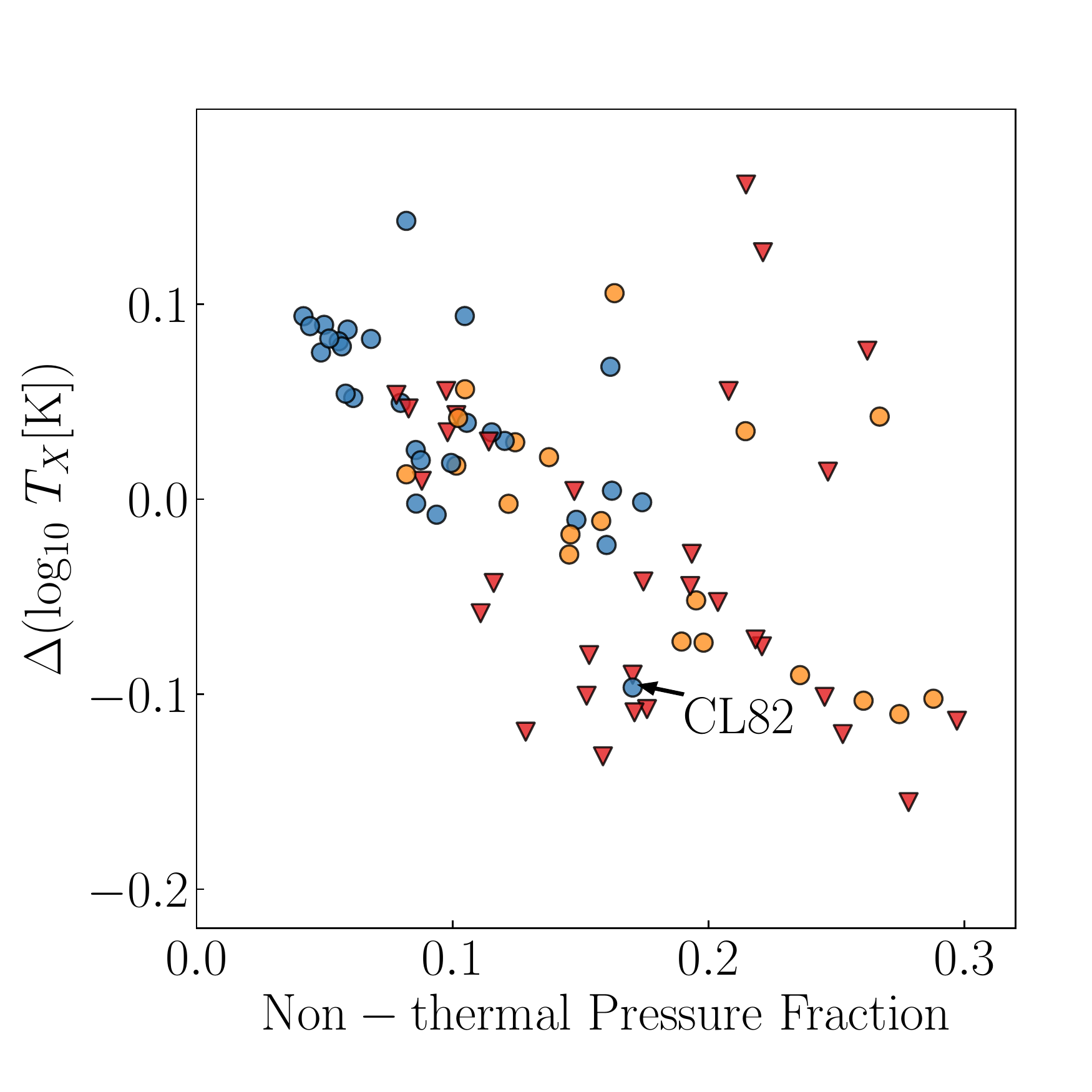}\hspace{-10pt}
  \includegraphics[width=0.343\linewidth]{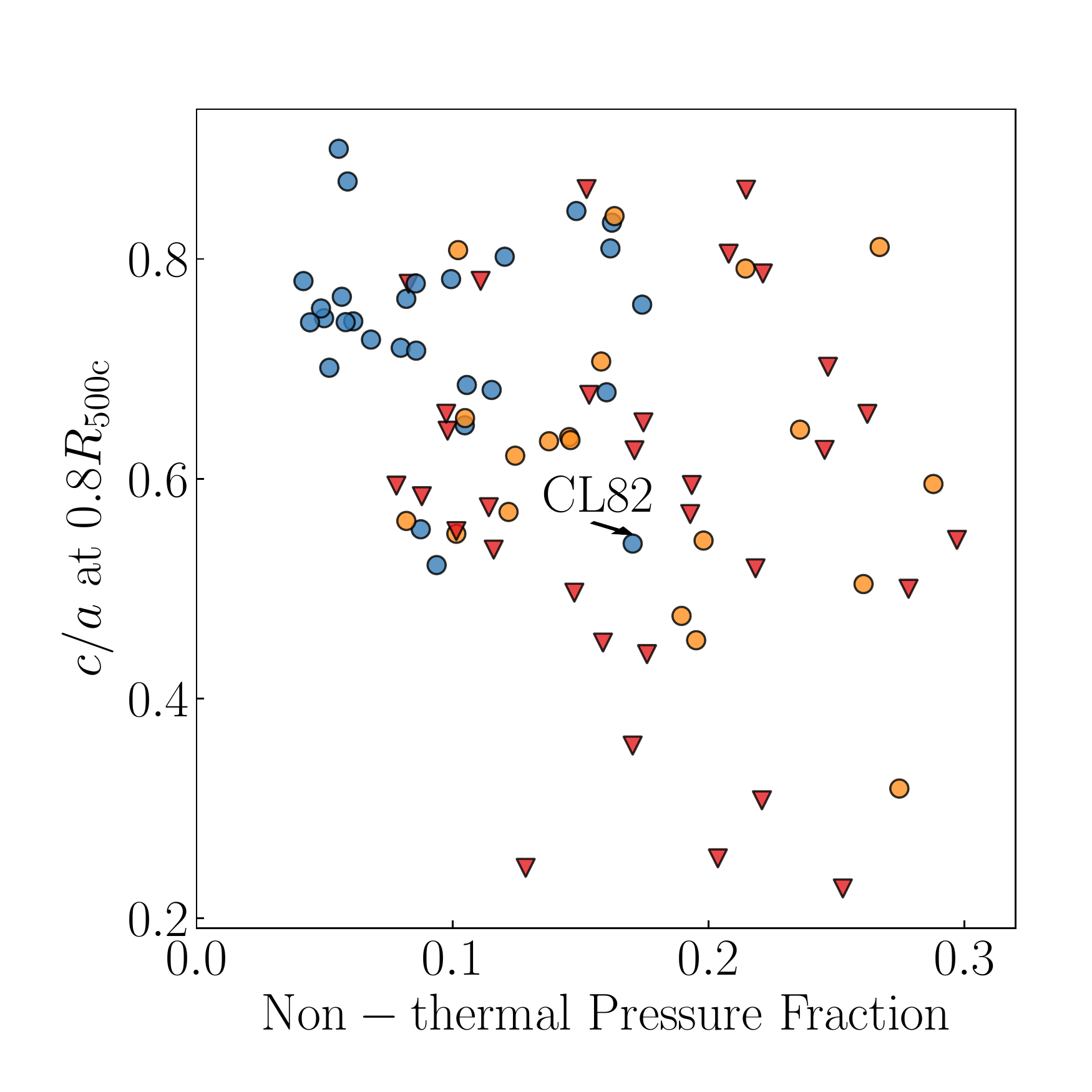}
\caption{Scatter plots of relations with different quantities, colour-coded by accretion modes the same as in Figure~\ref{fig.scatterplot} (red triangles -- major merger clusters, orange circles -- moderate merger clusters, blue circles -- smooth accreting clusters). Left: Non-thermal pressure fraction as a function of $\Gam{0.7}$. Middle: $\Delta \log_{\rm 10}(\Tx)$ as a function of non-thermal pressure. Note that the smoothly accreting cluster CL82 has a large non-thermal pressure fraction, similar to many major merger clusters. Right: Cluster ellipticity at $0.8 \r500c$ as a function of non-thermal pressure fraction. The correlation is weaker than in the left two figures because non-thermal pressure is an integrated quantity, while ellipticity is measured at a specific radius with a narrow annulus. Some major merger clusters with high non-thermal pressure fraction are in a stage of the merger when two cores collide, which results in a relatively round  outer region and weakens the correlation for shape and non-thermal pressure fraction.
}\vspace{10pt}
\label{fig:fnt}
\end{figure*}

%%%%%%%%%%%%%%%%%%%%%%%%%%%%
%%%%%%%%%%%%%%%%%%%%%%%%%%%
%%RESUME
%%%%%%%%%%%%%%%%%%%%%%%%
%%%%%%%%%%%%%%%%%%%%%%%%
% Checking thermalization hypothesis for relatively smooth correlation
To check the role of thermalization in residuals of the $\TxM$ relation, we compute the non-thermal pressure fraction, $\fnt$ of the clusters within $\r500c$.  The non-thermal pressure fraction is defined as,
$$\fnt = \frac{\Prand}{\Prand+\Pth}=\frac{\sigma^2_{\rm gas}}{\sigma^2_{\rm gas}+3k_{\rm B}T/\mu m_{\rm p}},$$
where $k_{\rm B}$ is the Boltzmann constant, $m_{\rm p}$ is the proton mass, and $\mu$ is the mean molecular weight. Both pressure terms $\Prand$ and $\Pth$ are mass weighted integrated quantities within $\r500c$ \citep[see][for details of the calculation of $\fnt$]{nelson14a}.

% Correlation between fnt, accretion rate, and residual (regardless of mode)
The left and middle panels of Figure~\ref{fig:fnt} show that there is a clear correlation between $\fnt$ and $\Gam{0.7}$ and residuals, $\Delta \Tx$. CL82 has a non-thermal pressure fraction of 17\%, which is larger than almost half of the major merger systems in our sample.  
% The non-thermal pressure fractions of major merger systems range between $0.085\lesssim \fnt\lesssim 0.3$.  
In contrast, the major merger system and moderate merger system, located in the low-$\Gam{0.7}$ positive $\Delta T$ part in the left-most panel of Figure~\ref{fig:fnt}, have only non-thermal pressure fractions of 8\% and 10\%, respectively.  We therefore conclude that the residual of the X-ray temperature from the best-fitting $\TxM$ relation is controlled by the overall MAR, rather than solely by merger events. 

% Check the correlation strength between bias in scatter and gamma.
The Spearman's rank coefficient of the correlation between $\Delta T_{\rm X}$ and $\Gam{0.7}$ is  $-0.55^{+0.10}_{-0.07}$.  The Spearman's rank coefficients for $\Delta T_X$ and $\Gam{0.75}$ and $\Gam{0.8}$ have  similar values of $-0.57^{+0.08}_{-0.08}$  and $-0.56^{+0.09}_{-0.09}$ , respectively.  The correlation is  weaker ($\sp > -0.48$) for time intervals larger or smaller than that. This is  similar to the time scale that drives the relation between the accretion rate and the ICM shape.

%  Transitive connection between Tx and shape
The fact that time intervals that maximize correlation between ellipticity and $\Tx$ residual with MAR are similar suggests that we can use ellipticity to select samples of rounder clusters to decrease scatter in the $\TxM$ relation. In other words, the ICM ellipticity can be used to select 
cluster samples with more relaxed and thermalized ICM which thus exhibits smaller scatter in the mass-observable relations.

% \ca{[I would put the rest of this paragraph and the next before you discuss thermalization.]}
Indeed, the upper right panel of Figure  \ref{fig:txscaling} shows that rounder clusters  mostly lie above the best-fitting power law relation, while  clusters with high ICM ellipticity lie mostly below. When we divide the sample into two equal populations defined by ellipticity there is a 0.07 dex difference in temperature between the two populations.
%  Direct correlation between Tx and shape
The lower right panel shows the residual of clusters from the best-fitting $\TxM$ relation as a function of 3D ICM ellipticity measured at $0.8\r500c$.  There is a clear correlation between temperature residual and $c/a$. 
 This continuous trend provides the groundwork for theoretical modeling of the intrinsic scatter of $\TxM$ relation, which is crucial for deriving true total mass and constraining cosmological parameters. We note a caveat that the 3D ICM ellipticity measurement is also affected by centroid-shift effects (important at inner radii) that decrease $c/a$, and gas from subclusters that have been removed (important at outer radii). Our ellipticity measurements are not equivalent to the ellipticity measured directly from X-ray isophotes, especially with projection effects.  Improvements to the calibration of the $\TxM$ relation should be accompanied by further detailed study of how to parameterize shape from mock X-ray images of individual surveys, as well as using simulations with more realistic galaxy formation physics.

%%%%%%%%%%%%%%%%%%%%%%%%%%%%%%%%%%%%% 
\section{Conclusion}
%%%%%%%%%%%%%%%%%%%%%%%%%%%%%%%%%%%%% 

We measure the shape of the intracluster medium in 80 massive clusters from the \Omegasim non-radiative cosmological simulation. We use the moment of inertia tensor method to self-consistently quantify cluster morphology and find the following:

\begin{itemize}
\item[1.] Our main result is that there is a correlation between ICM ellipticity at both inner and outer radii and cluster mass accretion rate (MAR). The correlation is strongest for MAR defined as the change of $\log_{10}\,M_{\rm 200m}$ between the expansion factor of $a_i=0.7$  and current epoch, $\Gam{0.7}$. There is almost no correlation of ellipticity with accretion rates defined with $\Gam{0.5}$ or $\Gam{0.9}$. Thus, our results show that accretion within a time-scale of $\approx 4.5$~Gyr (time between $\ai=0.7$ and $a_0=1.0$) has the strongest impact on the ICM morphology at $z=0$.  

\item[2.] The correlation between accretion rate and ellipticity at the outer radii holds for subsamples of all merger modes in the outer radii of $\geq 0.8\r500c$.  This indicates that the overall MAR, not just mode of accretion, determines the cluster ellipticity at $\geq 0.8\r500c$. On the other hand, we find that the ICM ellipticity at the inner radii is most sensitive to recent major mergers. 

\item[3.] 
Both the MAR and ellipticity correlate with the temperature residuals from the best-fitting power law $T_X-M$ relation, $\Delta T_X$. Notably, we find that there is a continuous systematic trend between cluster accretion rate and spectroscopic-like X-ray temperature, regardless of accretion mode.  Fast accreting clusters are cooler due to increased non-thermal gas motions from accretion.  The most relevant accretion time scale with the strongest $\Delta T_X-\Gamma$ correlation is $\sim 4$~Gyr, 
similar to the characteristic accretion time-scale that has the strongest correlation with ellipticity. Clusters above the median ellipticity (rounder) are 0.07 dex hotter than the clusters below the median ellipticity. Our results thus provide the basis for understanding and for theoretical modeling of the scaling relations, such as $\TxM$.
\end{itemize}

Our study shows that ellipticity of the ICM can be used as a probe or indicator of the MAR over the past $\sim 4$ Gyrs. 
We showed that, as a consequence of this, selecting clusters by ellipticity can lower the scatter of $\TxM$ relation.  
% Application to future surveys 
Future cluster surveys, derived from experiments such as {\it eRosita},  will have more than a thousand temperature measurements. The error in X-ray scaling relations will be dominated by astrophysical systematics \citep{hofmann17}. Our results show that some of these systematics can be mitigated by using the information encoded in the apparent ICM ellipticity. 

%%%%%%%%%%%%%%%%%%%%%%%%%%%%%%%%%%%%%%%%%%%%%%%%%%%%%%%%
\section*{Acknowledgments} 
The simulations were performed on the Omega HPC cluster at Yale. This work is supported in part by the facilities and staff of the Yale Center for Research Computing. CA acknowledges support from both the Enrico Fermi Institute and the Kavli Institute for Cosmological Physics at the University of Chicago through grant NSF PHY-1125897 and an endowment from the Kavli Foundation and its founder Fred Kavli.  The authors thank Philip Mansfield for useful discussions during this project.

\bibliographystyle{mnras}
\bibliography{ms} 

%%%%%%%%%%%%%%%%%%%%%%%%%%%%%%%%%%%%%%%%%%%%%%%%%%

%%%%%%%%%%%%%%%%% APPENDICES %%%%%%%%%%%%%%%%%%%%%

%%%%%%%%%%%%%%%%%%%%%%%%%%%%%%%%%%%%%%%%%%%%%%%%%%

% Don't change these lines
\bsp	% typesetting comment
\label{lastpage}
\end{document}